# Valley-hybridized gate-tunable 1D exciton confinement in MoSe$_2$

Maximilian Heithoff[1], Álvaro Moreno[1], Iacopo Torre[1], Matthew S. G. Feuer[2], Carola M. Purser[2,3], Gian Marcello Andolina[1], Giuseppe Calajo[1,4], Kenji Watanabe[5], Takashi Taniguchi[6], Dhiren Kara[2], Patrick Hays[7], Sefaattin Tongay[7], Vladimir Falko[8,9,10], Darrick Chang[1,11], Mete Atatüre[2], Antoine Reserbat-Plantey[1,12], Frank Koppens[1,11]

[1]ICFO – Institut de Ciencies Fotoniques, 08860 Castelldefels (Barcelona), Spain
[2]Cavendish Laboratory, University of Cambridge, JJ Thomson Avenue, Cambridge CB3 0HE, UK
[3]Cambridge Graphene Centre, University of Cambridge, UK
[4]Instituto Nazionale di Fisica Nucleare (INFN), Sezione di Padova, I-35131 Padova, Italy
[5]Research Center for Electronic and Optical Materials, National Institute for Materials Science, Tsukuba, Japan
[6] Research Center for Materials Nanoarchitectonics, National Institute for Materials Science, Tsukuba, Japan
[7]School for Engineering of Matter, Transport and Energy, Arizona State University, Tempe, United States
[8]National Graphene Institute, Manchester, UK
[9]Department of Physics and Astronomy, University of Manchester, UK
[10]Henry Royce Institute for Advanced Materials, Manchester, UK
[11]ICREA, Barcelona, Spain
[12]Université Côte d'Azur, CNRS, CRHEA. Valbonne, Sophia-Antipolis, France
*Corresponding authors: antoine.reserbat-plantey@cnrs.fr, frank.koppens@icfo.eu*

**Abstract**

Controlling excitons at the nanoscale in semiconductor materials represents a formidable challenge in quantum photonics and optoelectronics fields. Achieving this control holds excellent potential for unlocking strong exciton-exciton interaction regimes[1,2], enabling exciton-based logic operations[3], exploring exotic quantum phases of matter[4,5], facilitating deterministic positioning and tuning of quantum emitters[6,7], and designing advanced optoelectronic devices[8,9]. Monolayers of transition metal dichalcogenides (TMDs) offer inherent two-dimensional confinement and possess significant binding energies, making them promising candidates for achieving electric-field-based confinement of excitons without dissociation. While previous exciton engineering strategies have predominantly focused on local strain gradients[10], the recent emergence of electrically confined states in TMDs has paved the way for novel approaches[11]. Exploiting the valley degree of freedom associated with these confined states further broadens the prospects for exciton engineering. Here, we show electric control of light polarization emitted from one-dimensional (1D) quantum-confined states in MoSe$_2$. By employing non-uniform in-plane electric fields, we demonstrate the in-situ tuning of the trapping potential and reveal how gate-tunable valley-hybridization gives rise to linearly polarized emission from these localized states. Remarkably, the polarization of the localized states can be entirely engineered through either the spatial geometry of the 1D confinement potential or the application of an out-of-plane magnetic field.

**Introduction**

The manipulation of the dimensionality of a physical system has proven to be highly effective for controlling the behavior of quasiparticles. Prominent examples include the discovery of the quantum Hall effect, Luttinger liquids[12], electron quantum dots[13], hydrodynamic phonon transport[14], and van der Waals heterostructures. Among various quasiparticles, engineering the dimensionality of excitonic potentials plays a pivotal role in designing efficient exciton dissociation in photodetectors[8] and establishing the foundations for quantum technologies centered around localized excitons, such as single photon sources[6,7], qubit arrays[15], and exciton routers[3].

Approaches for manipulating excitons in two-dimensional (2D) semiconducting crystals can be broadly categorized into three primary classes. The first category involves the local trapping of excitons in zero-dimensional systems, thereby enabling single photon emission[16]. Deterministic methods employed in this approach encompass the utilization of nano-pillars[6,7] to induce local strain, defect implantation through techniques like He-FIB[17], and, more recently, utilizing moiré potentials[1,15,18]. These endeavors hold significant promise for applications in quantum technologies[19], including quantum key distribution, photon storage exploiting collective effects in arrays of quantum emitters, multiplexing indistinguishable photon sources, and solid-state quantum simulators. However, this approach currently lacks *in situ* control over the energies of quantum emitters, which hampers efficient manipulation of ensembles of such emitters.

The second avenue involves leveraging interlayer exciton states in van der Waals heterostructures similar to initial studies in double quantum well[20]. In this context, an out-of-plane electric field can effectively couple to the dipole of the interlayer exciton, yielding significant achievements such as quantum-confined Stark tuning[21,22], electrical control of exciton dynamics[23], and the realization of exciton-based transistors[3] or routers[24]. However, several challenges must be addressed, such as the fabrication of high-quality heterostructure devices, intricate mesoscopic reconstruction of the heterointerface[25], and the relatively slow recombination rates observed in indirect interlayer excitons, limiting the photoluminescence brightness.

The third approach encompasses directly manipulating the dielectric or electrostatic environment of a neutral intra-layer exciton. Initial efforts in this direction involved the creation of non-homogeneous dielectric environments by partially covering a TMD with graphene or hexagonal boron nitride (hBN)[26]. Such in-plane heterostructures lead to a significant alteration of electronic properties at the interface edge[11,26]. The precise nature of 1D excitonic states at these edges has been further explored in systems characterized by 1D strain gradients[10]. Furthermore, by harnessing the substantial in-plane electric field at the boundary between two ferroelectric domains[27], exciton drift and dissociation effects have been observed in monolayer TMD when deposited atop such domains. There has been considerable momentum in optoelectronics towards designing efficient photodetectors by engineering ultra-sharp PN junctions using split gates[8,9]. While these structures promote exciton dissociation, the large excitonic binding energy provides an opportunity for exciton confinement through electrostatic gates. A particularly elegant method has recently been demonstrated[11] by exploiting the capability to fabricate lateral P-i-N junctions with dimensions as small as 10 nm in $MoSe_2$.

Intra-layer exciton confinement in electrostatic traps encounters a fundamental challenge posed by an in-plane electric field, which can induce the dissociation of excitons. Our study addresses this issue by leveraging the substantial binding energy of intra-layer neutral excitons in $MoSe_2$. Specifically, we apply an inhomogeneous electric field on a nanoscale region to achieve center-of-mass localization and observe discrete quantized states. Furthermore, we investigate the possibilities of engineering the trap's geometry to establish an artificial 1D exciton channel. The resulting confinement potential exhibits remarkable localization, with exciton confinement below 5 nm and a location situated merely a few tens of nanometers away from a designer top gate electrode. We demonstrate that the localized states within this setup are linearly polarized, and we explain this with a model that incorporates significant inter-valley exchange interaction. The hybridization between valleys can be further tuned with an out-of-plane magnetic field that leads to purely circularly emitted light from the 1D-localized exciton states. Lastly, we provide an understanding of the interplay between the out-of-plane magnetic field and the in-plane electric field, which sheds light on the underlying mechanisms governing the formation of such a 1D confinement potential.

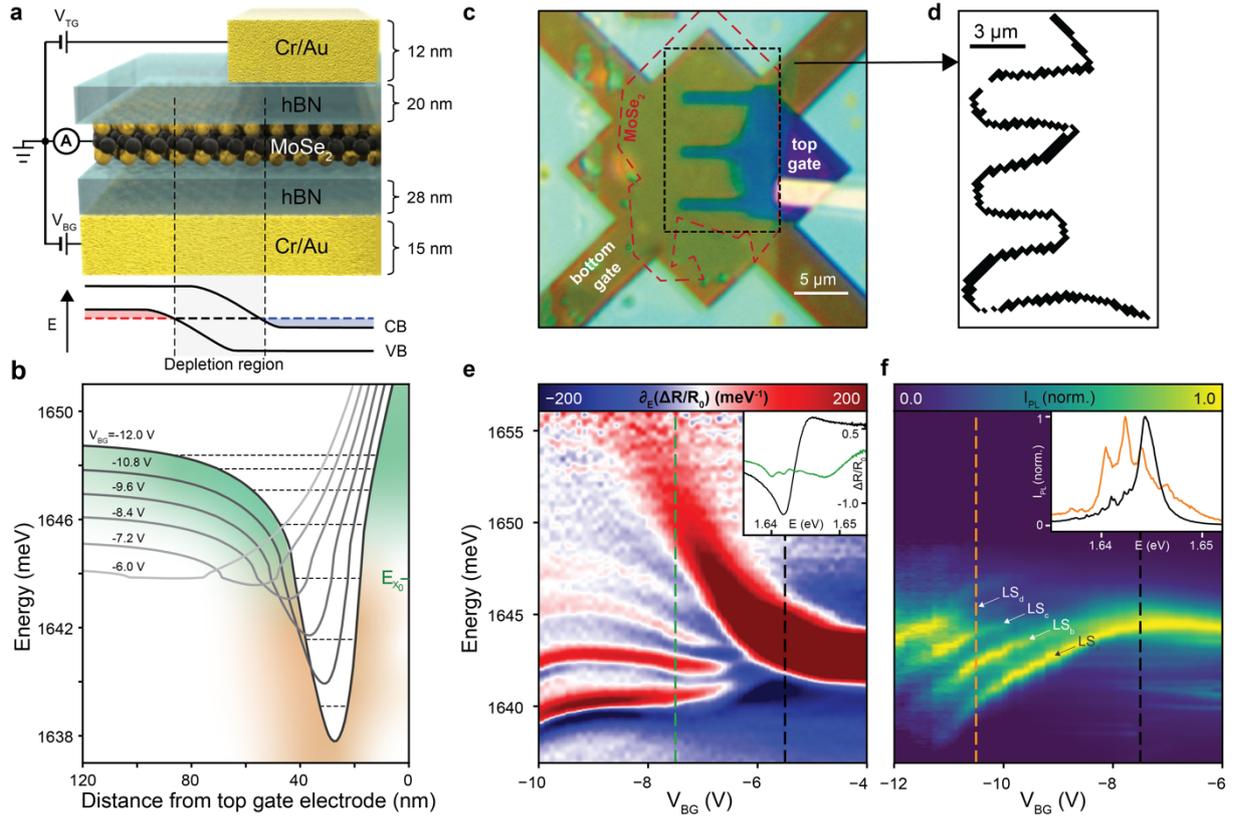

**Fig. 1: Gate-controlled 1D-excitonic potential in MoSe$_2$ monolayer. a:** Sketch of the device. This geometry unlocks high gradients in charge carrier densities and in-plane electric field at the top gate edge, where combination of Stark and interaction effects generates a narrow 1D-potential channel. P-i-N configuration around the top electrode is represented by the VB and CB edge diagram. **b:** Simulated trapping potential using Eq. 1 as a function of the in-plane distance from the top gate electrode edge for several values of $V_{BG}$. The green (orange) colour indicates the potential being mostly governed by interaction shift (Stark shift). Calculated localized states energy are shown (dashed lines) for $V_{BG} = -12$ V as well as the neutral exciton energy $E_{X_0}$ for comparison. **c:** Optical microscope image of the device. **d:** Modulus of the spatial gradient of differential reflectance taken at $E_{X_0}$ defined by $\nabla_{x,y}(\Delta R/R_0)|_{E_{X_0}}$. The objective is scanned in $(x,y)$ plane and $V_{BG} = -10\,V$. This quantity depicts the spatial onset of the 1D potential channel formed around the top gate electrode. **e:** Derivative of differential reflectance spectrum $\partial_E(\Delta R/R_0)$ versus $V_{BG}$. Single differential reflectance spectra taken at $V_{BG} = [-7.5, -5.5\,V]$ and represented by the vertical dashed lines (*cf.* respective colours in inset). **f:** Dependence of photoluminescence spectra with respect to $V_{BG}$. Each spectrum is normalized by its maximum. Inset shows two spectra taken at $V_{BG} = [-10.5, -7.5\,V]$. In **e,f** light is collected at the edge of the top gate electrode. All measurements and simulations are performed at $V_{TG} = 10\,V$ and $T = 3\,K$.

### Gate-tunable 1D confinement potential

The device consists of an hBN-encapsulated MoSe$_2$ monolayer placed in between a global bottom-gate and a local fingered top-gate (see Fig. 1a-b). The MoSe$_2$ monolayer covers most of the bottom-gate electrode, while top-gate fingers are 1 µm wide (see Fig. 1c). The bottom and top-gate potentials – $V_{BG}$ and $V_{TG}$, respectively – can be independently controlled while MoSe$_2$ is grounded. This asymmetric double gate geometry generates strong spatial inhomogeneities of charge carrier density $n$ within the MoSe$_2$ monolayer (see Fig. 1a-b). For instance, a negative bottom-gate potential ($V_{BG} = -10\,V$) leads to an excess of holes in the MoSe$_2$. By sufficiently increasing the top-gate potential ($V_{TG} = 10\,V$), negative charge carriers accumulate in the MoSe$_2$ area below the top-gate electrode. Such a gate combination creates a narrow PN-junction, with a depletion region's width governed by the large capacitive coupling between the top-gate and MoSe$_2$, facilitated by the thin top hBN spacer ($h_{hBN} \sim 20\,nm$)[11]. Only in this depletion region can the monolayer be efficiently penetrated by a large in-plane component of the electric field $\mathcal{E}$ generated at the top-gate edge, further enhanced by the small distance between the MoSe$_2$ and the top-gate electrode. Along the edge of the top-gate electrode, we can define a total exciton potential:

$$V_{tot} = -\frac{1}{2}\alpha \mathcal{E}^2 + \beta |n|, \qquad \text{Eq. 1}$$

with two contributions: i) within the narrow depletion region, a quadratic Stark potential originating from the large in-plane electric field $\mathcal{E}$ with polarizability $\alpha$, and ii) outside the depletion region, a repulsive interaction with scaling factor $\beta$ when excitons get dressed into polarons by the sea of free charge carriers[28,29]. This resulting 1D potential channel will lead to observable quantum-confined excitonic states if the energy splitting $\Delta E_n$ between states is larger than their linewidth. We reach that regime at cryogenic temperatures ($T = 3$ K), where encapsulated MoSe$_2$ usually exhibits narrow excitonic emission linewidths[30] $\Gamma \sim 1$ meV.

To monitor the emergence of this confinement potential, we perform low-temperature reflectance contrast (RC) and photoluminescence (PL) measurements as a function of gate voltages $V_{BG}$, $V_{TG}$ (see Fig. 1e-f). By setting the top-gate voltage at a high positive value ($V_{TG} = 10$ V), we can increase locally the density of negative charge carriers in MoSe$_2$ located below the top-gate electrode. When sweeping the bottom-gate voltage towards negative values, the hole density in MoSe$_2$ increases in the area uncovered by the top-gate electrode. Consequently, $V_{BG}$ acts as a knob to establish a confinement potential at the top-gate electrode edge (opposite polarity is shown in Supplementary Note 1). In Fig. 1e, we display RC as a function of $V_{BG}$ and plot its spectral differential for clarity. At $V_{BG} > -6$ V, we observe a stable and gate-voltage independent 1s neutral exciton line. There is a drastic behavior change around $V_{BG}^* = -6$ V: the neutral exciton strongly blueshifts, indicating an increasing p-doping, and four narrow additional lines appear with similar linewidth $\Gamma_{LS} \sim 1$ meV. The emergence of these additional lines in the spectrum can be attributed to localized states (temporarily referred to as $LS_{a-d}$) that are trapped within the 1D confinement potential created around the top-gate electrode. The spacing between consecutive states reaches a maximum value of $\Delta E_n \sim 2.4$ meV. This value suggests a tight center-of-mass confinement for the excitons with a typical spatial extent – within the harmonic oscillator approximation – defined by $\ell = \sqrt{\hbar/(m_X\omega)} \sim 5$ nm, with $m_X \sim 1.3\, m_e$ and $\hbar\omega = \Delta E_n$ being the energy splitting between localized states. The presence of redshifted and blueshifted localized states compared to the neutral exciton energy $E_{X^0}$ suggests the combined influence of both Stark and interaction-induced energy shifts on the formation of the trapping potential.

To further characterize these spectral features, we perform additional micro-photoluminescence (μ-PL) measurements (Fig. 1f) showing a similar fan of narrow lines for large negative bottom-gate voltages. The PL intensity of the neutral exciton line vanishes for $V_{BG} < V_{BG}^*$, as introducing additional charge carriers offers a decay pathway towards the energetically more favorable attractive polaron state[31]. For the localized states with a thermalized exciton population, we expect the lowest energy state – $LS_a$ – to feature the largest PL intensity. Interestingly, we observe a different behavior as the most intense peak is not always $LS_a$. For instance, the intensity of $LS_a$ is the highest for $V_{BG} > -10$V while $LS_b$ dominates around $V_{BG} \sim -10.5$ V. This is due to exciton population decay channels associated with exciton dissociation and polaron-exciton interactions[32]. Specifically, redshifted states compared to $E_{X^0}$ experience the influence of the in-plane electric field $\mathcal{E}$, which eventually becomes sufficiently strong to induce exciton dissociation.

Significantly, all the mechanisms described here strongly depend on the spatial position, as the confinement potential only appears near the top-gate electrode. To visualize this confinement potential, we record reflectance spectra for each point of the device and extract the spatial gradient of $\Delta R/R_0$ at the neutral exciton energy $E_{X_0}$, showing the spatial extend of the confinement potential (see Fig. 1d). Interestingly, the confinement potential strictly follows the edge of the top-gate electrode without discontinuities. In the following, we investigate the origin and geometry of this potential by performing polarization-resolved measurements.

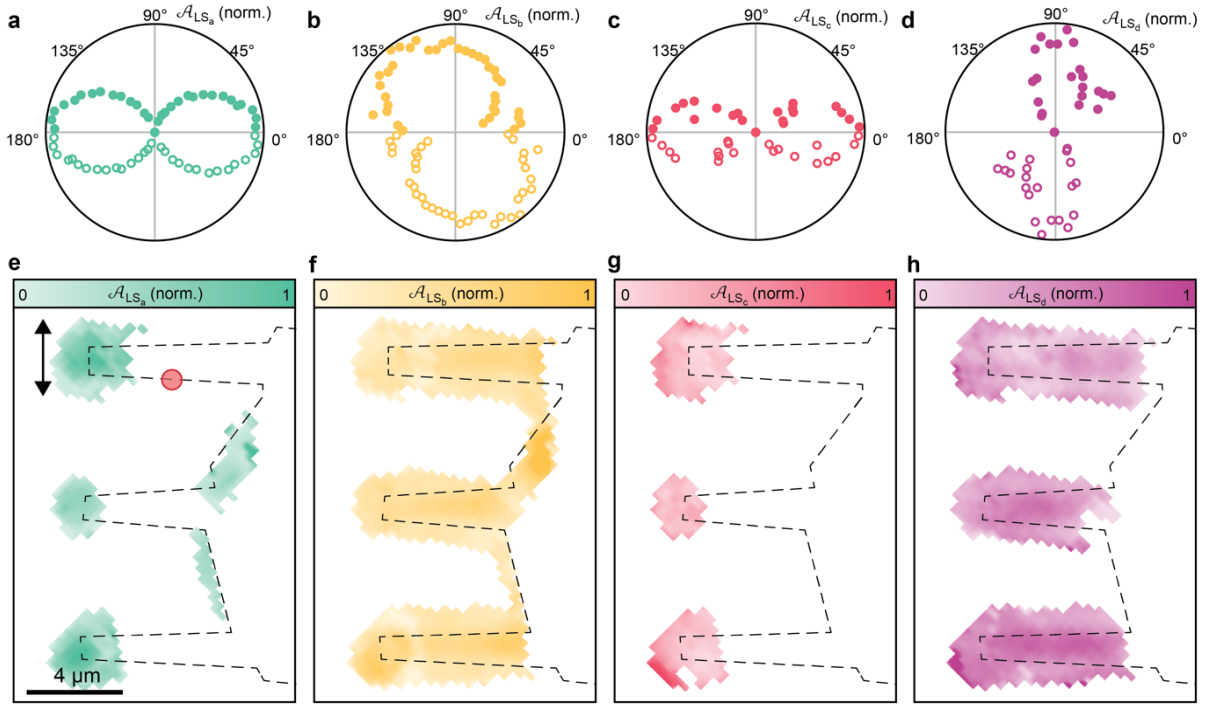

**Fig. 2: Polarized 1D-localized states in MoSe$_2$.** Fitted weight $\mathcal{A}$ of localized states' spectral components in RC as a function of the excitation polarization (**a-d**) and laser spot position (**e–h**). In all cases, both excitation and collection polarizations are colinear. In **a-d** the laser is focused at the edge of the top gate electrode (see red spot in panel **e**), while in **e–h** the excitation polarization is fixed is set at 90° (see arrow in panel **e**). For convenience in **a-d**, angle 0° corresponds to the longer edge of the top gate electrode fingers, with data from 0° to 180° mirrored. All measurements were taken at $(V_{TG}, V_{BG}) = (10, -9\ V)$, to maximise energy spacing of localized states $LS_{a-d}$, and at $T = 3K$.

**Signatures of linearly polarized 1D excitons**

We investigate the influence of the 1D confinement potential on the polarization of localized states by performing reflection measurements using linearly co-polarized excitation and detection. In monolayer TMDs, $K^+$ and $K^-$ valleys excitons can be selectively probed using $\sigma^+/\sigma^-$ light[33–35]. However, we expect the 1D excitonic confinement to perturb the valley circular dichroism. At the edge of the top gate finger, we record polarization-resolved differential reflection spectra, as shown in Fig. 2a–d. Each localized state is linearly polarized, and we find two groups ($LS_a$ ; $LS_c$) and ($LS_b$ ; $LS_d$), which are co-polarized either parallel (∥) or perpendicular (⊥) to the electrode's edge, respectively. The degree of polarization of $LS_b$ state is less pronounced and is discussed in the next section. These results imply that excitons near the top gate edge cannot be described in the circular-polarization-accessible base of $K$ and $K'$ valley but rather as a linear combination of both valleys – accessible via a pair of linearly polarized perpendicular light vectors. Such pairs of cross-polarized excitons have been reported on similar TMD monolayers.[11,36]

To track the polarization characteristics of the states along the edge of the top gate, we acquire hyperspectral reflectance maps with a fixed linear polarization shown in Fig. 2e. We observe distinct spectral weight patterns for $LS_{a-d}$, as depicted in Fig. 2e-h. Specifically, $LS_a$ and $LS_c$ are found exclusively at the vertical edges of the top gate's finger tips and along the diagonal sections that connect the fingers. Conversely, these states are absent along the longer horizontal edges of the fingers. In contrast, $LS_b$ and $LS_d$ emerge prominently along these longer edges of the fingers. These findings suggest that $LS_a$ and $LS_c$ exhibit parallel (∥) polarization with respect to the top gate edge, while $LS_b$ and $LS_d$ display perpendicular (⊥) polarization. Our study showcases the role of the top-gate electrode geometry in producing a strongly polarized 1D potential featuring localized excitonic states. Subsequently, we examine the underlying mechanism for generating cross-polarized pairs of localized excitons.

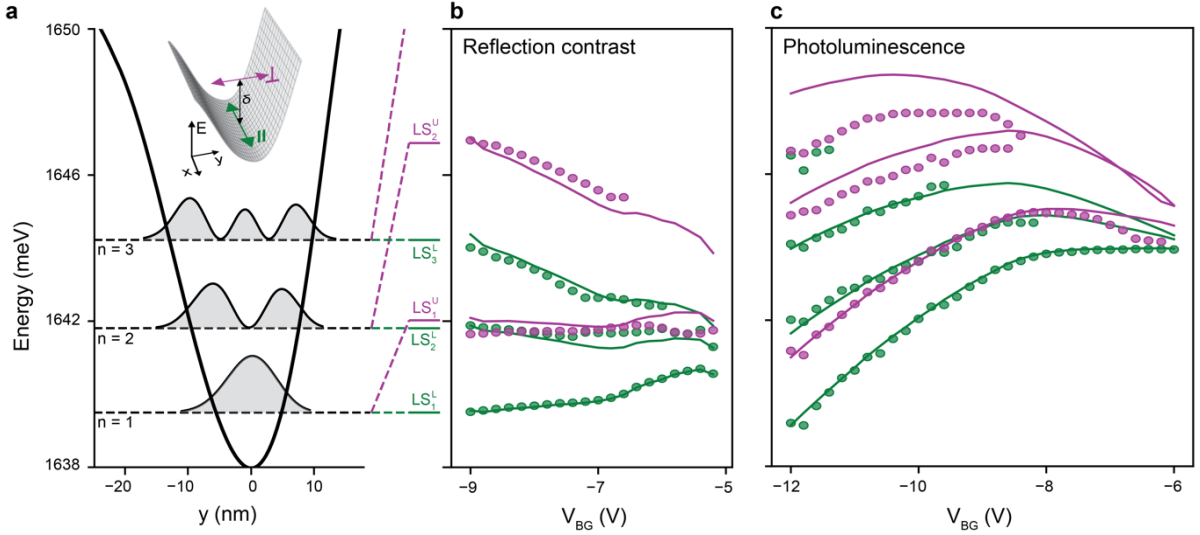

**Fig. 3: Enhanced valley exchange interaction in 1D trapping potential. a**: Simulated confinement potential using Eq. 1, showing quantized levels that coincide with the fitted results of the reflection data at $V_{BG} = -9\,V$. Such confined excitons show fine structure splitting, with lower/upper branches (green and purple dashed lines, respectively). These states are linearly polarized, respectively aligned either parallel ∥ or perpendicular ⊥ to the channel axis (inset). **b-c**: By adapting the exchange coupling constant $\gamma$, simulated (lines) states can be brought in close agreement with fitted energies of the localized states shown for ∥/⊥-polarization (green/purple dots, respectively). All reflection contrast (**b**) and photoluminescence (**c**) measurements taken at $T = 3$ K and $V_{TG} = 10$ V.

**Valley-hybridization of 1D-confined states**

We have established a clear correlation between the emergence of linearly polarized localized states and the qualitative description of an electrostatic 1D trapping potential. However, the precise formation and fine structure of these states remain elusive. Here, our approach involves simulating the accurate exciton potential near the top-gate edge. Our model considers the device geometry shown in Fig. 1a. By defining the gates' and monolayer electrochemical potentials and solving the system's non-linear Poisson equation (Supplementary Note 2), we obtain the spatial distribution of charge carrier density $n$ and in-plane electric field $\mathcal{E}$. For each input $(V_{TG}, V_{BG})$, we compute $n$ and $\mathcal{E}$, and extract the exciton potential using Eq. 1. By solving the Hamiltonian of the exciton's center-of-mass motion, we obtain multiple localized state solutions $LS_n$ (with $n$ being the index of the localized state). Every $LS_n$ can be adequately characterized as a standing wave, with an average wavevector denoted as $\bar{k}_n$. Notably, $\bar{k}_n$ lies beyond the boundaries of the light cone, exhibiting values approximately equal to 10 times the light wavevector ($q_{light}$) (Supplementary Note 3). In practical terms, this implies a substantial transfer of momentum from the trapping potential to the exciton, whereby $\bar{k}_n$ progressively increases as the trapping potential becomes stiffer (cf. Fig. 3a). This results in a fine structure splitting of each $LS_n$ into a lower branch $LS_n^L$ ∥-polarized and an upper branch $LS_n^U$ ⊥-polarized (inset Fig. 3a) with:

$$E_n^{U/L} = E_n + \frac{\delta_n}{2}(1 \pm 1). \qquad \text{Eq. 2}$$

The splitting is governed by long-range exchange interaction[37,38], which results in an exchange-interaction splitting $\delta_n = \gamma \bar{k}_n$ where $\gamma$ is the exchange-interaction coupling parameter. To resolve $\delta_n$, we record RC and PL spectra as a function of $V_{BG}$ under linear ∥/⊥ polarizations, as shown in Fig. 3b-c. Lower branches $LS_n^L$ can be resolved under ∥-polarization and give direct information about the trapping potential's energy levels as by definition $E_n^L = E_n$. To match simulated energy levels to this data, we set exciton masses to $m_{eff} = 1.3\ (1.1)\ m^*$ in RC (PL). On the other hand, upper branches $LS_n^U$ can be resolved under ⊥-polarization, and the exchange-interaction splitting is obtained from $\delta_n = E_n^U - E_n^L$. We show an agreement with the simulations for all energy levels over a large gate voltage range by matching a single constant coupling parameter $\gamma = 14\ (22)\ \text{meV} \cdot \text{nm}$ in RC (PL). Interestingly, $\gamma$ is linked to the radiative decay rate through expression[37] $\Gamma_{rad} = n_{hBN} E_n/(\hbar^2 c) \cdot \gamma$ with hBN refractive index $n_{hBN}$. This yields $\Gamma_{rad} = [0.39 - 0.61]\ \text{ps}^{-1}$ for $n_{hBN} = 2.18$ ([39]). Here, the trapping potential seems to slightly decrease recombination rates in comparison to the unbound exciton recombination rate of $\Gamma_0 \sim 3 \cdot \text{ps}^{-1}$, likely due to the in-plane electric field increasing electron-hole separation. Furthermore, the unbound exciton's Fourier limited linewidth sets a lower bound of $FWHM_{X_0} \geq \Gamma_0 \hbar/2 = 1.10\ \text{meV}$, while $FWHM_{LS} \geq [0.13 - 0.20]\ \text{meV}$ – consistent with the measured localized state linewidth as low as 0.5 meV. This combined analysis of gate-dependent measurements and simulations reveals that the previously identified $LS_b$ state consists of overlapping $LS_1^U$ (⊥) and $LS_2^L$ (∥) states. This observation demonstrates the presence of linearly polarized states $LS_{a,c,d}$ and partially polarized $LS_b$, which aligns with the data shown in the previous section (Fig. 2).

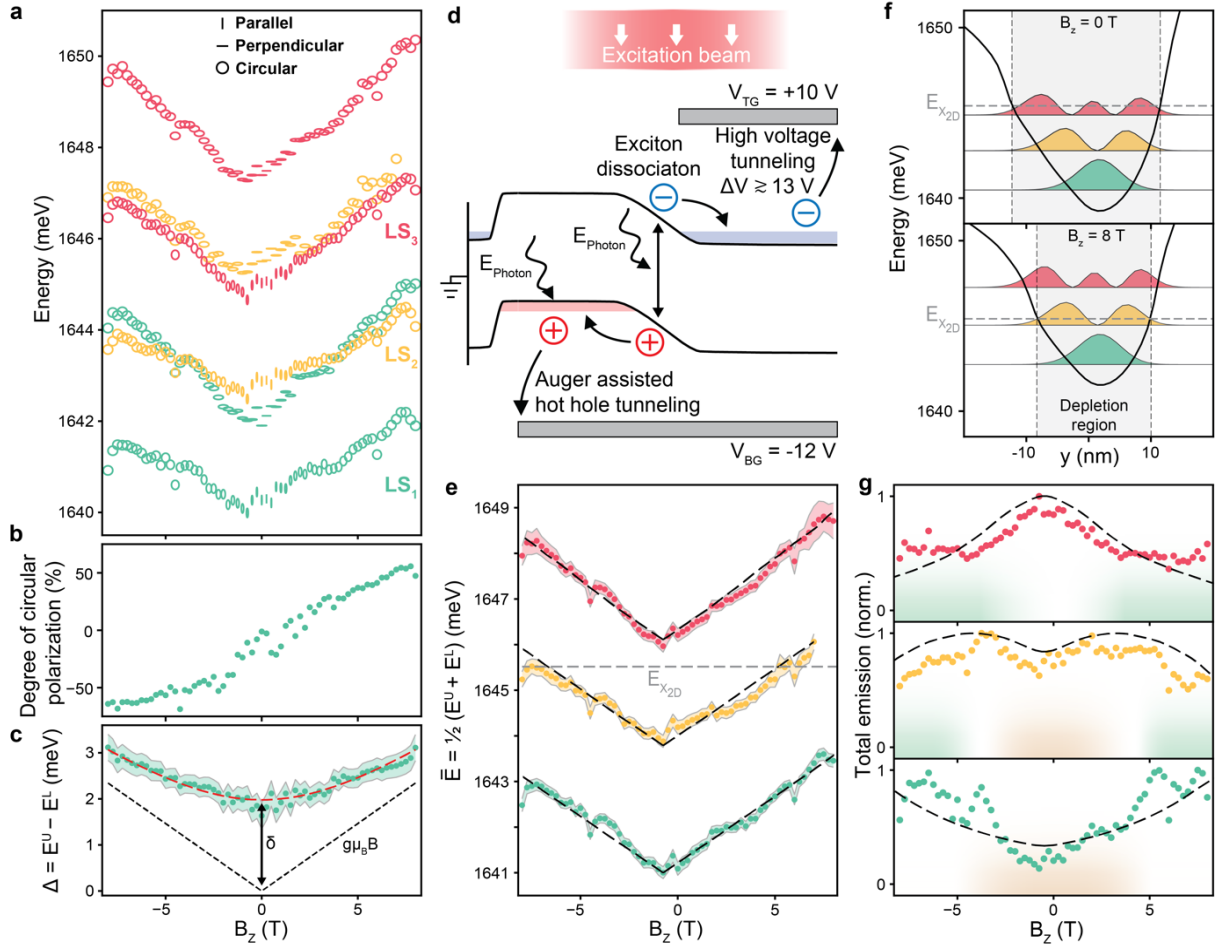

**Fig. 4: Photoluminescence of quantum-confined 1D exciton under out-of-plane magnetic field. a:** Localized state energy as a function of out-of-plane magnetic field $B_z$. For the three localized states identified by different colors, upper and lower branch energies $E^{U/L}$ are extracted. This data is compiled from four measurements under different detection polarizations (∥; ⊥; $\sigma^+$; $\sigma^-$) (Supplementary Note 4). In each of these measurements the excitation is $\sigma^+$ polarized. The degree of polarization is depicted by the shape of the symbol. **b:** Degree of circular polarization as a function of $B_z$ for $LS_1^L$ extracted from polarized PL measurements. **c:** Fine structure splitting $\Delta_1 = E_1^U - E_1^L$ as a function of the magnetic field and fitted data using Eq. 3 (red dashed line). **d:** Schematic band diagram at the top gate edge under illumination. Exact doping is given by the equilibrium between exciton dissociation and Auger assisted hot hole tunneling to the bottom-gate electrode. **e:** Average energies of $LS_{1,2,3}$ as a function of $B_z$. For each value of the magnetic field, we assign a charge carrier distribution and compute trapping potential as well as localized states energies (dashed lines) or wavefunctions (see cases for $B_z = 0; 8\,T$ shown in **f**). **g:** Total emission intensity for each localized state $LS_{1,2,3}$ as a function of $B_z$. Normalized exciton population (dashed lines) is calculated taking into account different decay channels associated to exciton dissociation (orange shading) and polaron-exciton interactions (green shading). All measurements are taken at $(V_{TG}; V_{BG}) = (10; -12\,V)$ and $T = 3\,K$.

## Magnetic modulation of trapping potential

While magneto-optics typically provide insights into the inner structure of the exciton, an out-of-plane magnetic field $B_z$ can also locally affect the electrostatic environment. Solving the $B_z$-dependent Hamiltonian of the localized exciton, we obtain a new expression of Eq. 2 for the upper ($U$) and lower ($L$) states energies:

$$E_n^{U/L} = E_n^X + \frac{\delta_n}{2} + E_{dia} \pm \frac{1}{2}\sqrt{\delta_n^2 + (g_n\mu_B B_z)^2} \qquad \text{Eq. 3}$$

where $E_{dia} = e^2 \langle r_n^2 \rangle B_z^2/(8\mu)$ is the diamagnetic shift with $\langle r^2 \rangle$ being the confined exciton root-mean-square (rms) size, and $\mu$ the exciton reduced effective mass. The Zeeman shift is $g_n \mu_B B_z$, where $g_n$ is the g-factor for localized exciton and $\mu_B$ is the Bohr magneton. In Fig. 4a, we show polarization-resolved PL measurements of localized states as a function of $B_z$. While the localized states are linearly polarized at $B_z = 0$ T (see Fig. 2), they become circularly polarized for $|B_z| > 0$. This effect is clear for the lower branch of $LS_1^L$, as shown in Fig. 4b, and is consistent with an increased lifting of valley degeneracy with $B_z$ (Supplementary Note 4). The fine structure splitting $\Delta_1 = E_1^U - E_1^L$ for $LS_1$ is shown in Fig. 4c. At $B_z = 0$ T, $LS_1^L$ and $LS_1^U$ are energetically separated by the exchange-interaction splitting $\delta$. For large magnetic fields, the Zeeman splitting dominates and the fine structure splitting becomes linear with $B_z$. Fitting the data with Eq. 3, we obtain $\delta_1 = 2.0$ meV and $g_1 = 5.1$, comparable with delocalized exciton g-factors in encapsulated[40] MoSe$_2$. The diamagnetic shift, typically associated with the exciton internal structure, is captured by the average energy of each localized state, $\bar{E}_n = (E_n^U + E_n^L)/2 = E_n^X + \delta_n/2 + E_{dia}$, as depicted in Fig. 4e. Remarkably, all localized states exhibit a significant linear shift with the magnetic field (> 230 μeV/T). This finding aligns with the previously reported linear dependence[11] of $\bar{E}_n(B_z)$. It is intriguing because the diamagnetic shift is expected to scale quadratically with $B_z$, considering that the magnetic length, $l_c = \sqrt{\hbar/eB_z} \sim$ 26 nm for $B_z = 1$ T, is substantially larger than the nanometer scale of the confined exciton[41].

To provide a potential explanation for this behavior, we develop a model based on a dependence of the charge carrier density on the magnetic field, as illustrated in Fig. 4d. Through the unique doping configuration, both the n-doped region under the top gate and the whole bottom-gated p-doped region is isolated from the electrical ground by depletion regions. We identify exciton dissociation at the illuminated top gate edge as an effective doping mechanism. The exact charge carrier distribution is then given by an equilibrium between dissociation and tunneling to the gate electrodes *via* electron tunneling[42] and Auger-assisted hot hole tunneling[43]. In the latter case, holes from $K^+/K^-$ valleys gain energy through Auger recombination. At such elevated energies, z-components of bands at the $\Gamma$ point hybridize with hBN bands, facilitating hole tunneling to the bottom-gate electrode. This process requires $K \to \Gamma$ phonon-assisted scattering. With increasing magnetic field, the valence bands at $K^+/K^-$ may become more resilient to such spin-depolarizing scattering events[44], thereby reducing the Auger tunneling rate.

To model this phenomenon, we assign a charge carrier distribution $n(B_z)$ for each magnetic field value by fitting the lowest state energy $\bar{E}_1(B_z)$. Subsequently, we calculate the trapping potential, energies, and wavefunctions of the localized states as illustrated in Fig. 4e-f. In this model, we implicitly hypothesize that the magnetic field induces modifications in the distribution of charge carriers, consequently altering the trapping potential. Experimental evidence supporting this model includes three independent observations. First, the measured photocurrent, corresponding to the tunneling towards the bottom-gate electrode, decreases by 25% when the magnetic field is switched to $B_z = 8$ T (data shown in Supplementary Note 5). Second, the linear dependence of the average energy $\bar{E}_n(B_z)$ of the confined states is accurately fitted by our model, as shown in Fig. 4e. This agreement additionally resolves the discrepancy discussed in ref.[11], between a considerable rms size of the localized exciton, extracted from a hypothetical diamagnetic shift ($\Delta \bar{E}_n \sim 2$ meV at 8 T), and the observed significant oscillator strength in RC measurements (cf. Fig 1e). The final evidence pertains to the change in total emission intensity, depicted in Fig. 4g. As the emission polarization changes with $B_z$, it is necessary to extract the total emission intensity, also known as the $S_0$ Stokes parameter, for each localized state (Supplementary Note 4). This enables us to distinguish intensity variations caused by the polarizers from the inherent decay of the exciton population and shows that the total emission of $LS_{1,2,3}$ varies with $B_z$. Specifically, as $B_z$ increases, the intensity of $LS_1$ decreases, displaying an opposite trend to that of $LS_3$. $LS_2$ is an intermediate situation, with a slight intensity increase at low magnetic fields and a reduction at larger values.

This behavior can be explained by the magnetic-field-dependent charge carrier distribution, which affects the strength of Stark and interaction confinement for each localized state (Supplementary Note 6). For instance, localized states like $LS_1$, which have lower energy compared to the unconfined exciton $X_{2D}$ (indicated by the dashed line in Fig. 4e), are primarily confined by the Stark effect (orange shading in Fig. 4g). On the other hand, higher energy states like $LS_3$ are strongly influenced by the interaction shift (green shading in Fig. 4g) as their wavefunction extends into regions with higher doping. The in-plane electric field decreases at higher magnetic fields, resulting in reduced exciton dissociation, thereby naturally increasing the $LS_1$ intensity. In contrast, $LS_3$ is more affected by the exciton-polaron interaction shift than Stark confinement. The polaron coupling is enhanced at larger $B_z$ values, creating a non-radiative decay channel that decreases emission intensity for $LS_3$. Our model is built upon the influence of an out-of-plane magnetic field, which alters the distribution of charge carriers in MoSe$_2$ near the top-gate edge. The precise underlying mechanism can be attributed to a reduction in spin-depolarizing scattering events[44] occurring within the valence bands at K$^+$/K$^-$ points. This reduction leads to a decrease in the rate of Auger tunneling as the magnitude of the magnetic field is increased. This model could be further tested with time-resolved valley polarization measurements.

**Discussion**

Our study presents an approach for achieving electrostatically quantum-confined intra-layer excitons in 1D channels. Three key aspects are highlighted: first, a large inhomogeneous in-plane electric field enables quantum confinement of neutral excitons. As initially demonstrated[11], the spatial extent of the localization potential is not limited by lithographic resolution, suggesting alternative methods, such as exploiting ferroelectric material grain boundaries[27]. In addition, confining intra-layer excitons offers enhanced light-matter interactions and the possibility of coupling to nanophotonic architectures, compared to inter-layer exciton manipulation. Second, gate-tunable valley hybridization facilitates electrically addressable 2D devices exploiting *in situ* tunable chiral light-matter interactions that can be further enhanced using the efficient coupling of 2D heterostructures with metasurfaces[45], for instance. Finally, the geometry of the top electrode dictates the 1D confinement potential, which can be further engineered[46] to create 0D structures[47] and periodic arrays made of quantum emitters[15] or valley-selective quantum emitters[48]. These configurations are promising for investigating quantum-collective effects and their applications in emergent quantum technologies, including quantum simulation[48], sensor arrays[49], and photon storage[50].

**Methods**

Reflectance contract (RC) measurements are performed in a confocal microscope setup with 0.7 NA (100X Mitutoyo Plan Apo NIR HR) and a closed-cycle cryostat (Montana Cryostation). For excitation, we use a supercontinuum laser source combined with a variable wavelength filter with a 6 nm spectral window (SuperK Extreme) with typical power of 10 nW and spot size ~750 nm. The polarization is controlled by a single linear polarizer on an automated rotation stage in front of the objective.

Micro-photoluminescence (µ-PL) measurements are performed in a confocal (SM/SM) microscope setup with 0.81 NA (Attocube LT-APO/NIR/0.81) within a closed-cycle cryostat (Attocube AttoDRY1000). We use a 660 nm cw-laser source (Laser Quantum Ventus 660) with 2 µW power and ~550 nm spot-size for excitation. Excitation and detection polarization are separately controlled by units of $\lambda/2$ and $\lambda/4$, each on automated rotation stages. Both setups use XYZ scanning stages.

Device preparation: layers (exfoliated MoSe$_2$, hBN, graphite) are picked up with a polypropylene-carbonate (PPC) coated PDMS stamp at 100°C for the hBN and graphite and room temperature for the TMD. The complete stack is dropped on a Si/SiO$_2$ (285nm) chip with pre-patterned electrodes for the bottom-gate and TMD, which is contacted with a few layers graphite flake. The release is done at 180°C, and the residual film is removed in chloroform and isopropanol. Lastly, the top gate is patterned directly on the stack by electron beam lithography and thin film evaporation deposition. All the contacts are made of 2nm Ti layer and an Au layer of 15nm for the pre-patterned electrodes, and 10nm for the top gate.

**Acknowledgments**

The authors thank Evgeny Alexeev, François Dubin, and Andreas Stier for constructive discussions and Matteo Ceccanti for his advice and help in the sample gates fabrication. F.H.L.K. acknowledges support from the ERC TOPONANOP (726001), PCI2021-122020-2A funded by MCIN/AEI/ 10.13039/501100011033), the "European Union NextGenerationEU/PRTR (PRTR-C17.I1), D.E.C acknowledges support from the European Union, under European Research Council grant agreement No 101002107 (NEWSPIN). F.H.L.K and D.E.C acknowledge support from the government of Spain (PID2019-106875GB-I00; Severo Ochoa CEX2019-000910-S [MCIN/ AEI/10.13039/501100011033], Fundació Cellex, Fundació Mir-Puig, and Generalitat de Catalunya (CERCA, AGAUR, 2021 SGR 01443). Authors thank European Union's Horizon 2020 under grant agreement no. 881603 (Graphene flagship Core3) and 820378 (Quantum flagship). A.R-P thanks support from UCAJEDI ANR-15-IDEX-01 and Doeblin FR 2800. G. C. thanks European Union Horizon 2020 research and innovation programme under the Marie Sklodowska-Curie grant agreement No 882536 (QUANLUX) and QUANTERA 2021 (T-NiSQ). M.S.G.F. thanks EPSRC Doctoral Training Programme. C.M.P. thanks ERC Advanced Grant PEDESTAL (884745). K.W. and T.T. acknowledge support from the JSPS KAKENHI (Grant Numbers 21H05233 and 23H02052) and World Premier International Research Center Initiative (WPI), MEXT, Japan. S.T acknowledges support from DOE-SC0020653, NSF CMMI 1825594, NSF ECCS 2052527, DMR 2111812 and CMMI 2129412.

# Supplementary Information:

## Table of contents

## Supplementary Note 1: Appearance of localized states under alternative top and bottom gate configurations

In addition to the RC data shown in Fig. 1e of the main text we perform further measurements under alternative gate configurations. In Fig. S1a, we fix the bottom instead of the top gate voltage at $V_{BG} = -10\text{ V}$ and sweep the top gate voltage. At $(V_{BG}\,;\,V_{TG}) = (10\,;-10\text{ V})$, we intersect with the voltage sweep of Fig. 1e. Consequently, we observe the same set of localized states as described in the main text, showing similar behaviour in the voltage range $V_{TG} \in [6; 10\text{V}]$. In contrast, Fig. S1b shows a different behavior. Due to the reversed voltage configuration of top and bottom gate P and N doped regions of the monolayer are switched. According to Fig. 4d of the main text, this results in the N doped region being electrically connected to ground preventing from and increased accumulation of charge carriers in this region. This allows the electrical field to penetrate the monolayer more effectively, promoting an increased Stark effect. In the voltage range $V_{BG} \in [0; 10\text{ V}]$, we observe localized states redshifted by up to $10\text{ meV}$, while states shown in the PN-configuration are shifted only by a few meV.

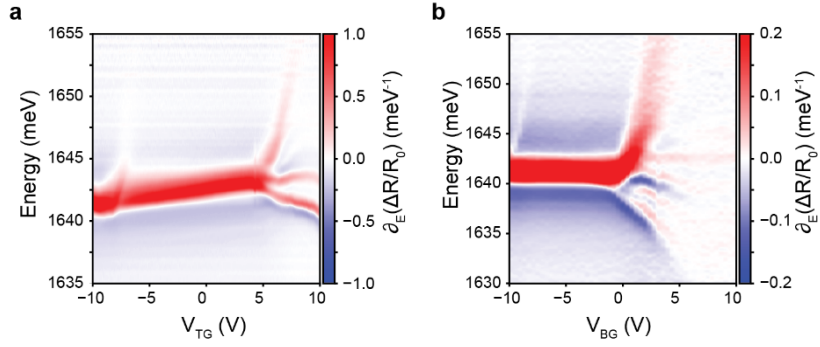

**Fig. S1: a:** RC data obtained for $V_{BG} = -10\,V$ while varying $V_{TG}$ to show the equivalence of tuning either the top or bottom gate to create the confining potential. In both cases, the laser spot was focused at the edge of the top gate in similar experimental condition as in Fig. 1 of the main text. Narrow features depicting localized states appears at $V_{TG} > 6V$. **b:** RC data obtained for $V_{TG} = -11.8\,V$ while varying $V_{BG}$. This dataset shows that localized states are visible also in the "NP-junction" case for $V_{BG} > 0V$. This situation is opposite to the "PN-junction" case presented in the main article.

## Supplementary Note 2: Electrostatic simulations of the exciton trapping potential

To simulate the exciton trapping potential around the top-gate electrode edge, we solve the non-linear Poisson equation using routines of the FEniCS problem solving environment in Python. Consequently, we can determine the exciton potential using Eq. 1: $V_{tot} = -\frac{1}{2}\alpha \mathcal{E}^2 + \beta|n|$ and the localized state by solving:

$$E_n\,\psi_n \;=\; H\,\psi_n \;=\; -\frac{\hbar^2}{2m_x}\,\partial_y^2\,\psi_n + V_{tot}\,\psi_n.$$

The resulting exciton potential is plotted in Fig. S2 for varying $V_{BG}$ and spatial position, where the top gate edge is located at 200 nm. Due to the PN-junction configuration of the device, as shown in Fig. 4d of the main text, the device behaves locally around the top gate finger as if it is not connected to ground. This allows us to adapt the charge carrier distribution in the MoSe$_2$ as an additional knob in the simulation. For each ($V_{TG}$; $V_{BG}$), we can find a charge carrier distribution for which the simulated lowest energy level $E_{LS_1}$ matches with the measured data.

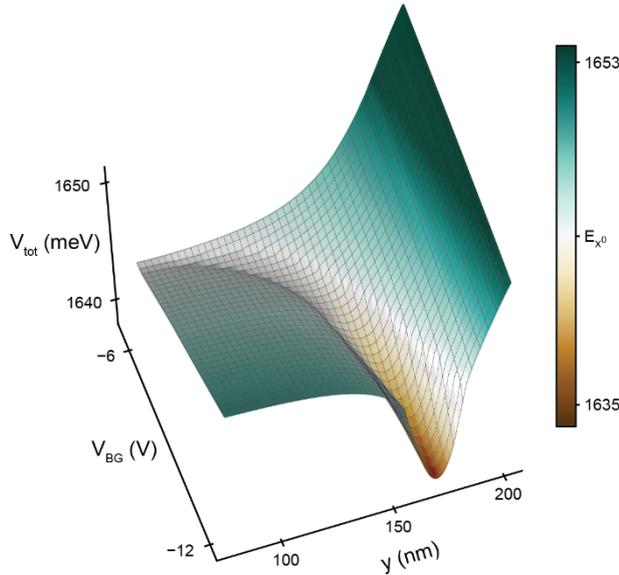

**Fig. S2**: Complementary exciton potential $V_{tot}$ for PL data in Fig. 3c of the main text simulated as a function of the applied bottom-gate potential ($V_{BG}$) and the position along the y-axis (in-plane with the MoSe$_2$ monolayer) normal to the electrode edge. The edge of the electrode corresponds here to $y = 200\,nm$. The color indicates the dominance of Stark effect (brown) and interaction shift (turquoise). For the simulation we fix $V_{TG} = 10\,V$. As in the publication from Thureja et al.[1] we assume an exciton polarizability $\alpha = 6.5\,eV\,nm^2\,V^{-2}$ and an dielectric polarizability of slab $hBN$-$MoSe_2$-$hBN$ $\epsilon_\parallel = 6.93$, $\epsilon_\perp = 3.76$ and a comparable exciton mass $m_x = 1.1$. We obtain overlapping results for measured and simulated localized states shown in Fig. 3c for an exciton-electron coupling strength $\beta = 0.25\,\mu eV\,\mu m^2$.

## Supplementary Note 3: Derivation of the valley exchange coupling parameter $\gamma$

In complement to Fig. 3a of the main text, Fig. S3 qualitatively showcases the stark contrast in k-vector of unbound (red dashed lines) and localized exciton states (grey dashed lines). For localized states, the intersection with the exciton dispersion shows a considerable exchange interaction splitting $\delta$, while splitting of the unbound exciton is much smaller than the measured exciton linewidth.

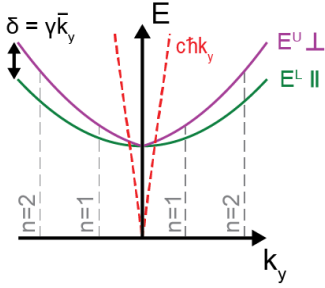

**Fig. S3:** Schematic illustration of the exciton dispersion showcasing the division of the bright doublet into eigenstates that are linearly polarized. These eigenstates have dipole moments aligned either parallel (L) or perpendicular (T) to the wavevector within the plane. Exact values of the average wave vector $\bar{k}_n$ can be extracted from the simulated wave functions: $\bar{k}_n = \langle \psi_n ||k|| \psi_n \rangle$. The diagram also includes red dashed lines indicating the light cone dispersion with much smaller wave vector k at exciton energies of $E_{X_0} = 1644\ meV$.

The valley exchange coupling scales with the exciton radiative decay rate. We use Eq. 12 from Ref[2] (rewritten in our notation) with radiative recombination rate $\Gamma_0$, average localized state momentum $\bar{k}_n$ and photon momentum $q = nE_n/(\hbar c)$ (refractive index $n$ and localized state energy $E_n$). In the limit of $\bar{k}_n \gg q$, we obtain:

$$\delta_n = \hbar\Gamma_0 \frac{\bar{k}_n^2}{q\sqrt{\bar{k}_n^2 - q^2}} \approx \hbar\Gamma_0 \frac{\bar{k}_n}{q} = \frac{\hbar^2 c \Gamma_0}{nE_n} \cdot \bar{k}_n.$$

Since $\delta_n = \gamma \cdot \bar{k}_n$, we obtain $\gamma = \hbar^2 c \Gamma_0 / (nE_n)$ and:

$$\Gamma_0 = \frac{nE_n}{\hbar^2 c} \cdot \gamma.$$

## Supplementary Note 4: Localized states' energies and emission intensities in B-field dependent measurements

Here, we want to resolve the energy and emission intensity of each localized state $LS_n^{U/L}$ with varying out-of-plane magnetic field $B_z$. As shown in Fig. 3 of the main text, the upper $LS_n^U$ and lower $LS_{n+1}^L$ branches largely overlap in energy due to the comparable magnitude of energy level splitting $\Delta E_{n,n+1}$ and exchange interaction splitting $\delta_n$. However, the emission from each $LS_n^{U/L}$ is fully polarized and each overlapping set of $LS_n^U$ or $LS_{n+1}^L$ can be described by mostly orthogonal vectors on the $S_1, S_3$ plane of the Poincaré sphere with $\theta_n^L = arctan\left(\frac{g_n \mu_B B_z}{\delta_n}\right)$ and $\theta_{n+1}^U = arctan\left(\frac{g_{n+1} \mu_B B_z}{\delta_{n+1}}\right)$ (Fig. S4) and are therefore distinguishable.

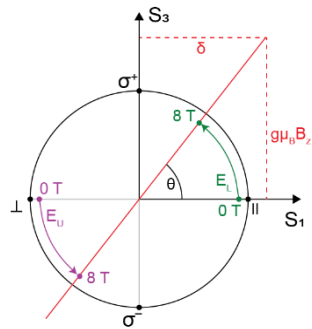

**Fig. S4: Cross section through the Poincaré sphere.** The 1D confinement potential introduces a zero-field magnetic splitting $\delta$ that splits each energy level into two branches ($\parallel, \perp$), polarized parallel and perpendicular to the top-gate edge.

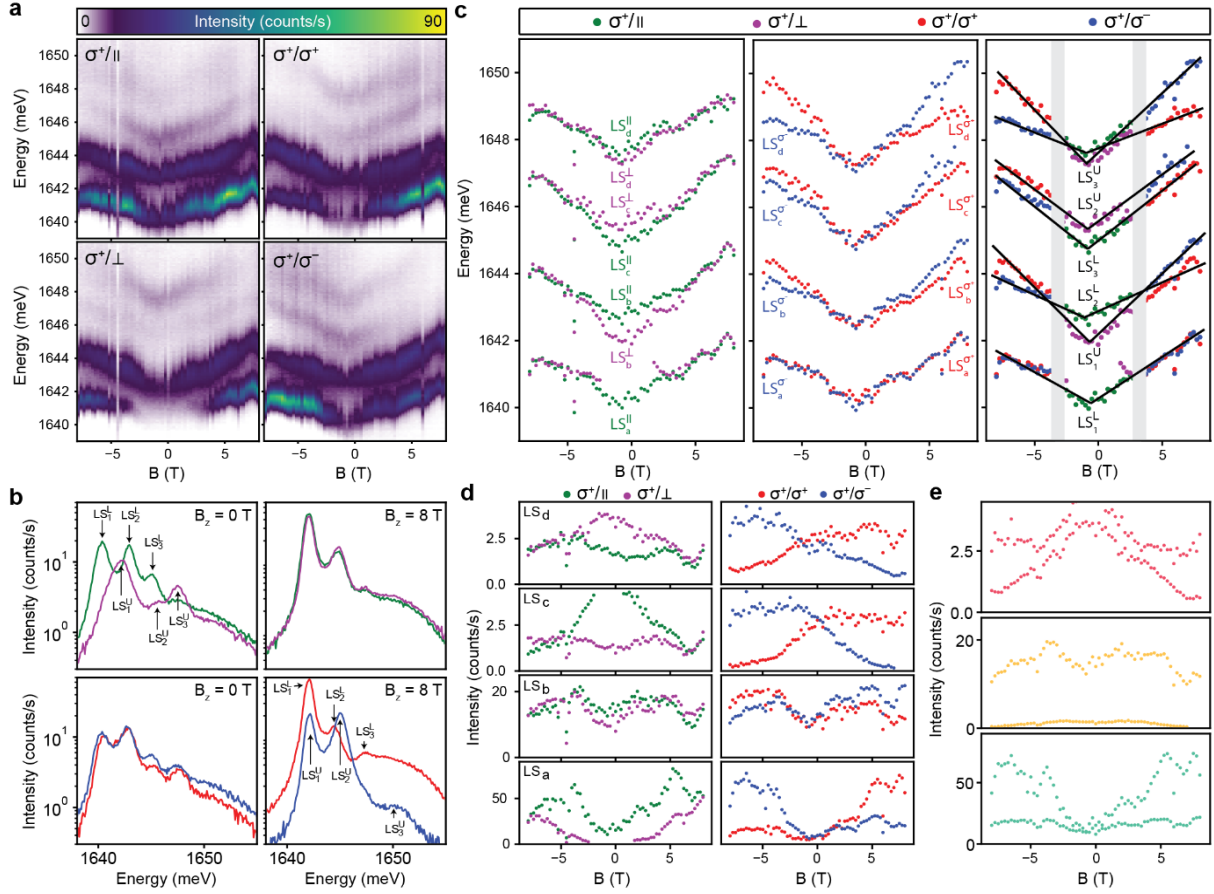

**Fig. S5: Polarization resolved magneto-photoluminescence measurements.** For all measurements the excitation at 660 nm is kept at $\sigma^+$ polarization, while we chose the polarization of the detection between linear polarized parallel $\parallel$ (green lines and dots) and perpendicular (magenta lines and dots) $\perp$ to the top-gate edge, as well as circularly polarized positively $\sigma^+$ (red lines and dots) and negatively $\sigma^-$ (blue lines and dots): **a:** Photoluminescence spectra as function of out-of-plane magnetic field $B_z$ taken at $(V_{TG}; V_{BG}) = (10; -12\,V)$ and $T = 4\,K$. **b:** Line cuts at 0 and 8 T for all four detection polarizations. Due to the distinct polarization of upper and lower branch of each energy level, we can resolve $LS_n^{L/U}(0\,T)$ in the $\parallel/\perp$ base (upper right panel) and $LS_n^{L/U}(8\,T)$ in the $\sigma^+/\sigma^-$ base (lower right panel) **c:** Fitted energies for all four polarizations for emission peaks $LS_{a-d}$ shown in two separate panels for better distinction. The same data is presented again in the third panel in order to demonstrate the synthesis of the energies of the localized states energies $E_{1-3}^{U/L}$ from $LS_{a-d}$. **d:** Fitted emission intensities of $LS_{a-d}$ shown in separate plots for better distinction. **e:** Synthesised emission energies for $LS_{1,2,3}^{U/L}$.

We measure the projection of such multi-polarized emission on the Stokes vectors $S_1$, $S_3$ by detecting with $(\parallel; \perp; \sigma^+; \sigma^-)$ polarization while maintaining the excitation $\sigma^+$ polarized (Fig. S5a). The dependence of the polarization base on $B_z$ is well illustrated in $\perp$ detection. The lowest energy peak is invisible at $0\,T$ as it is fully $\parallel$ polarized, but becomes visible under $\perp$ detection for $B_z \neq 0\,T$ due to its increasing circular polarization. Spectra at 0 T and 8 T are shown in Fig. S5b. Due to the fully linear polarization ($\theta_{n,n+1} = [90°, 270°]$) at 0 T, only $LS_n^L$ are visible under $\parallel$ detection and $LS_n^U$ under $\perp$ detection (upper left panel of Fig. S5b.), while measurements under circular detection show nearly no distinction. Likewise, at 8 T, measurements in linear detection show no linearly polarized features, while circular detection shows distinct peaks in $\sigma^+$ and $\sigma^-$ detection ($\theta_{n,n+1} \sim [0°, 180°]$). The projection of the localized states' spectral distribution $I_{LS_n^{U/L}}(\lambda)$ on linear and circular base is given by

$$\begin{pmatrix} I_\parallel(\lambda) \\ I_\perp(\lambda) \end{pmatrix} = \begin{pmatrix} \cos^2\theta_{n+1}/2 & \sin^2\theta_n/2 \\ \sin^2\theta_{n+1}/2 & \cos^2\theta_n/2 \end{pmatrix} \begin{pmatrix} I_{LS_{n+1}^L}(\lambda) \\ I_{LS_n^U}(\lambda) \end{pmatrix}$$

$$\begin{pmatrix} I_{\sigma^+}(\lambda) \\ I_{\sigma^-}(\lambda) \end{pmatrix} = \begin{pmatrix} \sin^2 \theta_{n+1}/2 + \pi/4 & \cos^2 \theta_n/2 + \pi/4 \\ \cos^2 \theta_{n+1}/2 + \pi/4 & \sin^2 \theta_n/2 + \pi/4 \end{pmatrix} \begin{pmatrix} I_{LS_{n+1}^L}(\lambda) \\ I_{LS_n^U}(\lambda) \end{pmatrix}$$

In the vicinity of two overlapping peaks $LS_n^U$ and $LS_{n+1}^L$ we could reconstruct the localized states' spectral distribution $I_{LS_n^{U/L}}(\lambda)$ from the measurements in $(\|;\perp;\sigma^+;\sigma^-)$ by inverting this mapping. However, as the exact values for $\theta_n^L$ and $\theta_{n+1}^U$ are unknown, we resort to fit measurements under each four polarizations using up to four pseudo-Voigt distributions $LS_{a-d}^{\|;\perp;\sigma^+;\sigma^-}$ extracting peak position $E_{a-d}^{\|;\perp;\sigma^+;\sigma^-}$ and peak intensity $I_{a-d}^{\|;\perp;\sigma^+;\sigma^-}$.

### a. Determination of LS energies

Knowing that the peaks are fully linearly polarized at $0\,T$ and nearly circularly polarized at $8\,T$, we identify $E_{1,2,3}^{L,U}(0\,T)$ localized states energies at $B = 0\,\text{T}$ by measuring emission in the $\|, \perp$ basis. Similarly, we identify $E_{1,2,3}^{L,U}(8\,T)$ measuring emission in the $\sigma^+, \sigma^-$. In order to be coherent with the initial identification of the localized peaks, we define $E_{1,2,3}^{L,U}(0\,T) = E_{a-d}^{\|,\perp}$ and $E_{1,2,3}^{L,U}(8\,T) = E_{a-d}^{\sigma^+,\sigma^-}$ (in Fig. S5c). For intermediate fields, we interpolate localized states energies from the weighted mean:

$$E_{1,2,3}^{L;U}(B_z) = \left(1 - \frac{B_z}{8\,T}\right) \cdot E_{a-d}^{\|;\perp}(B_z) + \frac{B_z}{8\,T} \cdot E_{a-d}^{\sigma^+;\sigma^-}(B_z)$$

For instance, for $LS_1$ we use following interpolation:

$$E_1^L(B_z < 0\,T) = \left(1 - \frac{B_z}{8\,T}\right) \cdot E_a^{\|}(B_z) + \frac{B_z}{8\,T} \cdot E_a^{\sigma^-}(B_z)$$

$$E_1^L(B_z > 0\,T) = \left(1 - \frac{B_z}{8\,T}\right) \cdot E_a^{\|}(B_z) + \frac{B_z}{8\,T} \cdot E_a^{\sigma^+}(B_z)$$

$$E_1^U(B_z < 0\,T) = \left(1 - \frac{B_z}{8\,T}\right) \cdot E_b^{\perp}(B_z) + \frac{B_z}{8\,T} \cdot E_b^{\sigma^+}(B_z)$$

$$E_1^U(B_z > 0\,T) = \left(1 - \frac{B_z}{8\,T}\right) \cdot E_b^{\perp}(B_z) + \frac{B_z}{8\,T} \cdot E_b^{\sigma^-}(B_z)$$

### b. Determination of LS intensities

We use the same approach to determine the intensities. Knowing that the peaks are fully linearly polarized at $0\,T$ and nearly circularly polarized at $8\,T$, we identify $I_{1,2,3}^{L,U}(0\,T)$ localized states emission intensity at $B = 0\,\text{T}$ by measuring emission in the $\|, \perp$ basis. Similarly, we identify $I_{1,2,3}^{L,U}(8\,T)$ measuring emission in the $\sigma^+, \sigma^-$. In order to be coherent with the initial identification of the localized peaks, we define $I_{1,2,3}^{L,U}(0\,T) = I_{a-d}^{\|,\perp}$ and $I_{1,2,3}^{L,U}(8\,T) = I_{a-d}^{\sigma^+,\sigma^-}$ (in Fig. S5d). For intermediate fields, we interpolate localized states intensity from the weighted mean:

$$I_{1,2,3}^{L;U}(B_z) = \left(1 - \frac{B_z}{8\,T}\right) \cdot I_{a-d}^{\|;\perp}(B_z) + \frac{B_z}{8\,T} \cdot I_{a-d}^{\sigma^+;\sigma^-}(B_z)$$

For instance, for $LS_1$ we use following interpolation:

$$I_1^L(B_z < 0\,T) = \left(1 - \frac{B_z}{8\,T}\right) \cdot I_a^{\|}(B_z) + \frac{B_z}{8\,T} \cdot I_a^{\sigma^-}(B_z)$$

$$I_1^L(B_z > 0\,T) = \left(1 - \frac{B_z}{8\,T}\right) \cdot I_a^{\|}(B_z) + \frac{B_z}{8\,T} \cdot I_a^{\sigma^+}(B_z)$$

$$I_1^U(B_z < 0\,T) = \left(1 - \frac{B_z}{8\,T}\right) \cdot I_b^{\perp}(B_z) + \frac{B_z}{8\,T} \cdot I_b^{\sigma^+}(B_z)$$

$$I_1^U(B_z > 0\,T) = \left(1 - \frac{B_z}{8\,T}\right) \cdot I_b^{\perp}(B_z) + \frac{B_z}{8\,T} \cdot I_b^{\sigma^-}(B_z)$$

## Supplementary Note 5: Photocurrent measurement under magnetic field.

In complementation to the model presented in Fig. 4d of the main text, Fig. S6 shows an electric current measured between top and bottom gate electrode upon illuminating the edge ($P_{\lambda=660\,nm} = 2\,\mu W$). Importantly, the current quenches without illumination, or when illuminating other parts of the sample. This illustrates how exciton dissociation and tunnelling currents act as the essential doping mechanisms of the semiconducting $MoSe_2$ monolayer. At $V_{BG} = -12$ V, we obtain the same experimental condition as for Fig. 4 of the main text. For an applied out-of-plane magnetic field $B_z = 8\,T$, the electric current and therefore the exciton dissociation reduces by $\sim 20\,\%$ indicating a reduced in-plane electric field $\mathcal{E}$ around the edge.

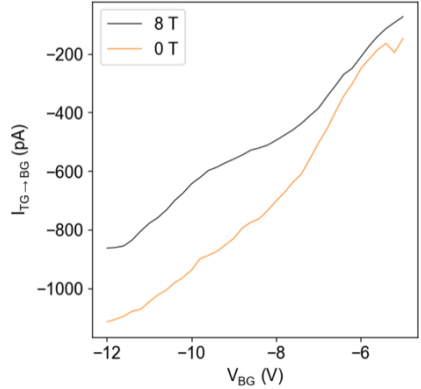

**Fig. S6:** *Measured photo-current from top-gate to bottom-gate electrode as a function of the applied bottom-gate potential.* The measurement is taken in the same conditions as is Fig. 4 of the main manuscript. The experiment is repeated for $B_z = 0\,T$ and $B_z = 8\,T$. At high magnetic field, the photocurrent is reduced due to less exciton dissociation, as argued in the main manuscript.

## Supplementary Note 6: Tuning exciton decay rates under a magnetic field

Fig. S7 shows complementing data to the simulated charge carrier distribution $n(B_z)$ and its effect on the emission intensity, shown in Fig. 4e-g of the main text. As described in the main text, for each value of $B_z$, we find $n(B_z)$ to match measured and simulated states' energies. Furthermore, from the simulations, we obtain an in-plane electric field distribution $\mathcal{E}(B_z)$ and for each localized state a wavefunction $\psi_n(B_z)$. Consequently, we calculate the average electric field $\langle\psi_n|\mathcal{E}|\psi_n\rangle$ (Fig. S7a) and charge carrier distribution $\langle\psi_n|n|\psi_n\rangle$ (Fig. S7b) felt by each localized state $LS_{1,2,3}$ at each value of $B_z$. In Figure S7c, we translate the average electric field into an exciton dissociation rate $\Gamma_{dissociation}$ according to numerical calculations[3] and the average charge carrier distribution into a decay rate into attractive polarons $\Gamma_{exciton-polaron} = \sigma\,\langle\psi|n|\psi\rangle$ with an empirically found exciton-charge carrier cross-section $\sigma$. Assuming a constant radiative decay rate $\Gamma_{radiative}(B_z)$, the exciton population can be expressed as $N_x \propto \left(\Gamma_{radiative} + \Gamma_{dissociation} + \Gamma_{exciton-polaron}\right)^{-1}$. The simulated results for the normalized exciton population for $LS_{1,2,3}$, shown in Fig. 4g of the main manuscript (dashed lines), are obtained using $\Gamma_{radiative} = 1\,\text{ps}^{-1}$ and $\sigma = 250\,\text{cm}^2.\text{s}^{-1}$.

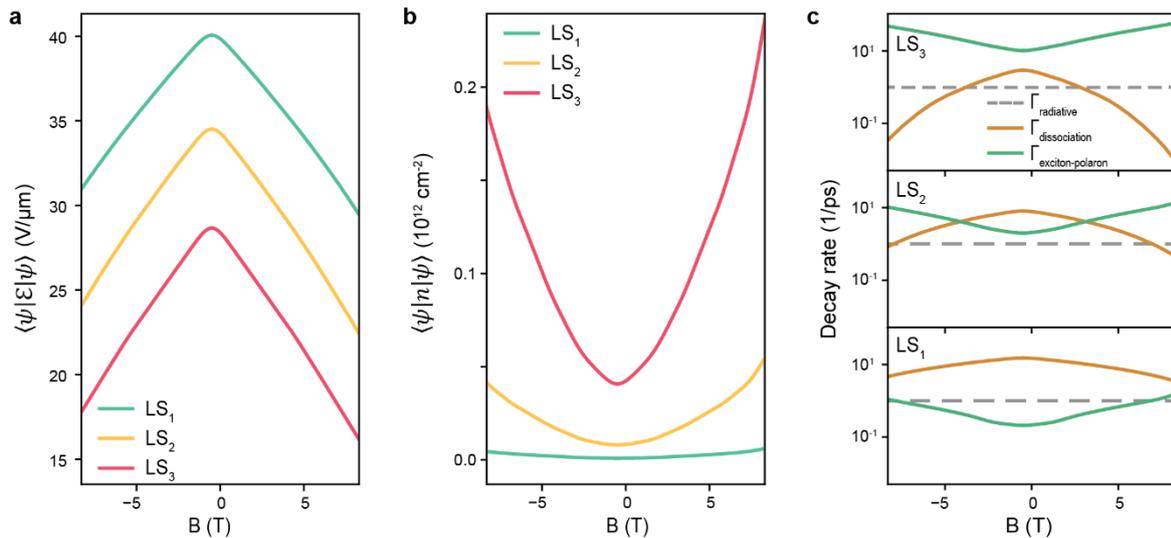

**Fig. S7: Determining decay rates from electrostatic simulations.** The localized states' wavefunctions $\psi$ exhibit varying overlap with the in-plane electric field $\mathcal{E}$ and charge carrier distribution $n$. Lowest energy $LS_1$ is mostly bound by Stark effect and experiences a stronger electric field, while higher energy $LS_3$ is mostly bound by the repulsive interaction with charge carriers as their wavefunction shows a stronger overlap with charge carriers of the p- and n-doped regions (c.f. Fig. 4f of the main text). **a,b:** Due to the magnetic field dependence of the p-doped region, we gain control over the average electric field $\langle\psi|\mathcal{E}|\psi\rangle$ and charge carrier distribution $\langle\psi|n|\psi\rangle$ felt by each localised state $LS_{1,2,3}$. **c:** localized state decay rate (log scale) as a function of the out-of-plane magnetic field.